\RequirePackage{iftex}
\ifpdftex
  \documentclass[reprint, aps,prl,superscriptaddress,floatfix,longbibliography]{revtex4-2}
  \usepackage{graphicx}
\else
  \documentclass[reprint, aps,prl,superscriptaddress,floatfix,longbibliography,dvipdfmx]{revtex4-2}
  \usepackage[dvipdfmx]{graphicx}
\fi
\usepackage{amsmath,amssymb,bm,color}
\usepackage[colorlinks=true,allcolors=blue]{hyperref}

\newcommand{\aMnTe}{$\alpha$-MnTe}

\newcommand{\Ueff}{U_{\mathrm{eff}}}

\begin{document}

\title{Magnetic reconstruction of the altermagnet
\aMnTe(0001) surface driven by ligand holes}

\author{Tomonori Tanaka}
 \email{tanaka.t.da74@m.isct.ac.jp}
\author{Yoshihiro Gohda}
 \email{gohda@mct.isct.ac.jp}
\affiliation{Department of Materials Science and Engineering, Institute of Science Tokyo, Yokohama 226-8501, Japan}
\date{\today}

\begin{abstract}
We show from first principles that the altermagnet \aMnTe{} reconstructs its
magnetic order at the Te-terminated (0001) surface. The ground state has a
ferromagnetic outermost Mn bilayer in place of the bulk-continued stacking. The
driver is the holes that the surface leaves on the Te mediating the exchange
within that bilayer. The computed constant-energy contours agree with
photoemission maps of the surface metal. Filling the holes restores the bulk order: the
carrier density can control the strength of the surface coupling and, by
reversing its sign, the surface magnetic order itself.
\end{abstract}
\maketitle
\textit{Introduction}---Altermagnetism is a collinear magnetic phase in
which two antiparallel spin
sublattices are related by a rotation rather than by translation or
inversion~\cite{Smejkal2022a,Smejkal2022b}. The bands then carry a
momentum-dependent spin splitting at essentially zero net magnetization, so
the magnet supports, without producing stray fields, transport responses
ordinarily reserved for ferromagnets~\cite{Smejkal2020,Gonzalez2023}. A prototype is the
semiconductor \aMnTe{} (NiAs structure, $P6_3/mmc$), whose Mn$^{2+}$ ions
($S=5/2$) order into an A-type
antiferromagnet: ferromagnetic (0001) planes stacked antiferromagnetically
along the $c$ axis~\cite{Kunitomi1964,Kriegner2016}. That order splits the magnon bands by chirality and the
electron bands by spin without spin--orbit
coupling~\cite{Liu2024,LiuChiral2026,Alaei2025,Krempasky2024} and, with
spin--orbit coupling, gives a crystal Hall
effect~\cite{Gonzalez2023}: an anomalous Hall conductivity that is
odd under reversal of the N\'eel vector and depends on its in-plane
orientation. It thus reads out an order parameter that carries no net moment
and that a magnetic field reaches only through spin--orbit-induced
canting~\cite{Jungwirth2016,Baltz2018}. In recent films the bulk
valence-band maximum lies below the Fermi level and the conduction is
independent of thickness beyond a few unit cells: it belongs to metallic
surface states of the Te-terminated (0001) face inside the bulk
gap~\cite{Zhou2026,SurfaceAHE2026}. A Hall measurement on such a film reads
the magnetic order of the outermost layers.

Those metallic surface states have a chemical origin, and that origin fixes
their filling. Electron counting for the polar Te
termination~\cite{Pashley1989,Harrison1979} leaves each surface Te short by
one electron: one hole per $1\times1$ surface, an integer carrier density
fixed by stoichiometry rather than by doping. The hole sits
on the Te: \aMnTe{} is a charge-transfer insulator, its ligand-to-$3d$
charge-transfer energy lying below the $3d$--$3d$ Coulomb
repulsion~\cite{Sato1995}, so a hole introduced into the crystal resides on
the ligand rather than changing the Mn valence~\cite{Martuza2025}. The Te in turn mediates
the interlayer exchange that fixes the A-type
stacking, and that exchange is short-ranged. A
surface seldom reverses the stacking of the bulk order, and in the known
cases the driver is an undercoordinated magnetic site: the terminal Mn of CaMnO$_3$(001) couples
ferromagnetically to the layer below~\cite{Filippetti1999}, and the surface
moments of Fe$_{1+x}$Te cant under relaxation~\cite{Trainer2021}. The
closest analogue of what follows is Cr$_2$O$_3$(0001), whose outermost Cr
bilayer orders ferromagnetically with the terminating O layer polarized
against it~\cite{Cline2000}. A hole on
the ligand would be a driver of a different kind, and there is precedent for
its sign: in $p$-type dilute magnetic semiconductors valence-band holes
couple $3d$ moments ferromagnetically against an antiferromagnetic
background~\cite{Ferrand2001,Dietl2014}. Whether the hole reverses the interlayer
bond its Te mediates is a quantitative question. To our knowledge the magnetic
ground state of this surface has not been determined, in experiment or in
calculation; the slab calculations to date take the bulk stacking for
granted~\cite{Zhou2026,SurfaceAHE2026}.

In this Letter, we show from first principles that the outermost Mn bilayer
of the surface is ferromagnetic, against an antiferromagnetic
interior: an enumeration of the interlayer spin configurations of
Te-terminated slabs finds the ground state at this reconstruction rather
than at the bulk-continued A-type order. Removing the carriers
electronically restores the bulk order; they drive the reversal through a
channel that closes when the anion shell is filled. The constant-energy
contours computed for the reconstructed surface are consistent with
published photoemission maps~\cite{Zhou2026}. Across the
outermost interlayer bond, then, the reconstruction breaks the sublattice
relation that
defines the altermagnetic phase, and leaves each surface with an uncompensated
moment.

\begin{figure*}
\includegraphics[width=\textwidth]{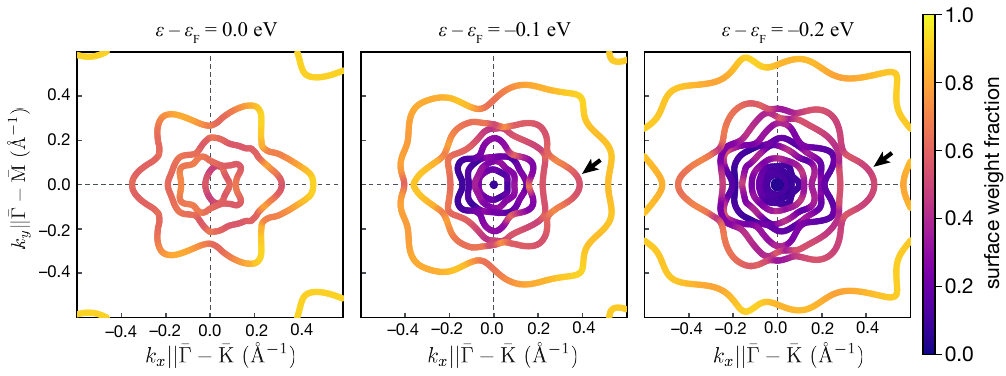}
\caption{Constant-energy contours of the reconstructed \aMnTe(0001) surface
at $\varepsilon-\varepsilon_\mathrm{F}=0$, $-0.1$ and $-0.2$~eV (left to
right). Every band crossing the energy is drawn, colored by the
fraction of its site-projected weight in the outermost four planes (two
Te and two Mn) of the metallic face: bright sheets are surface-localized,
darker ones carry their weight in the deeper planes of the slab. The axes follow the zone: $k_x$ runs along
$\bar\Gamma$--$\bar{\mathrm{K}}$ and $k_y$ along
$\bar\Gamma$--$\bar{\mathrm{M}}$. The common direction of the two surface Mn
moments is $+k_y$; the N\'eel axis of the interior is parallel to them.
Arrows mark the surface sheet whose petals lie along
$\bar\Gamma$--$\bar{\mathrm{K}}$ at $-0.1$ and $-0.2$~eV, the sheet
compared with photoemission in the text; on the bulk-continued surface the
corresponding petals lie along $\bar\Gamma$--$\bar{\mathrm{M}}$. That map,
and the full-zone view of both stackings, are in the Supplemental
Material~\cite{SM}.}
\label{fig:fs}
\end{figure*}

\textit{Methods}---We used density functional theory within
the projector augmented-wave method~\cite{Blochl1994} as implemented
in
VASP~\cite{Kresse1996,Kresse1999}, with the PBE exchange-correlation
functional~\cite{Perdew1996} and an on-site Coulomb correction on the Mn $3d$
states in the Dudarev scheme, $\Ueff=4.0$~eV~\cite{Dudarev1998},
the value used in prior bulk exchange studies of \aMnTe~\cite{Alaei2025}.
Plane waves were expanded to a $450$~eV cutoff and the surface Brillouin zone
sampled on a $\Gamma$-centered $12\times12\times1$ mesh per $1\times1$ cell,
with total energies converged to $10^{-6}$~eV. Spin--orbit coupling~\cite{Steiner2016} is
included in every calculation reported; only the
geometries were relaxed without it. Slabs are symmetric and Te-terminated ($13$ planes,
carrying six Mn planes, for the enumeration; $17$ and $21$ planes for the
thickness check;
$15$~\AA{} vacuum), built at the lattice constants of the bulk relaxed in the
A-type state at the same level of theory ($a=4.21$~\AA, $c=6.76$~\AA). The Te
termination is favored over the Mn termination everywhere in the accessible
chemical-potential window~\cite{SM}. The
in-plane cell is $1\times1$ throughout: cleaned single crystals diffract as a
sharp $1\times1$ with no superstructure~\cite{Martuza2025}. For the
band structure of Fig.~\ref{fig:bands} and the constant-energy contours of
Fig.~\ref{fig:fs}, one face is passivated
with fractionally charged pseudo-hydrogens along its severed bonds
($20$~\AA{} vacuum, with a dipole correction for the now asymmetric
cell~\cite{Neugebauer1992,Bengtsson1999}) so that a single metallic face
remains~\cite{SM}. We
enumerated all $32$ inequivalent collinear interlayer spin configurations of
the 6-Mn slab, computed their static energies, and fitted them to
a layer-resolved spin model $E=E_0-\sum_{l<l'}J_{ll'}\sigma_l\sigma_{l'}$,
with $\sigma_l=\pm1$ the orientation of the ferromagnetically ordered $l$th Mn
plane numbered from the surface, $E_0$ a configuration-independent constant,
and $J>0$ favoring parallel alignment.
Reconstruction energies are quoted per $1\times1$ surface (half the
difference for the two-faced slab), the unit that carries one ligand hole.
Full methodology and the robustness checks are in the Supplemental
Material~\cite{SM}.

\begin{figure}
\includegraphics[width=\linewidth]{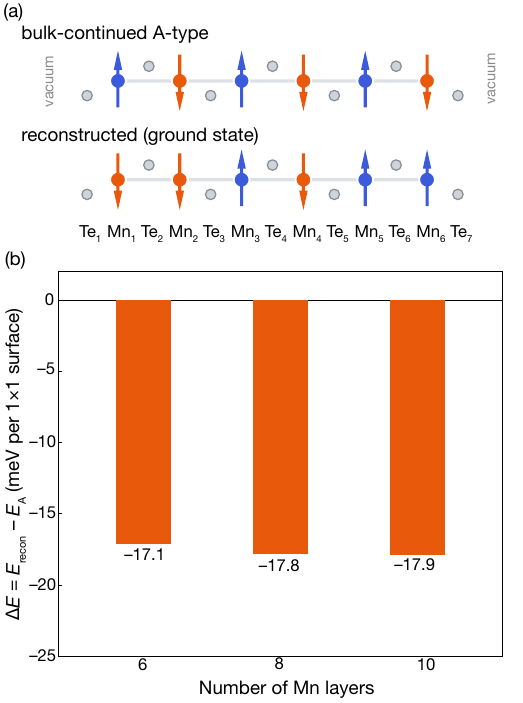}
\caption{Surface magnetic reconstruction. (a)~The two competing states of
the 6-Mn-layer slab, drawn sideways with the Mn moments in the basal plane: the
bulk-continued A-type stacking, and the reconstructed ground state, which
reverses the outermost Mn plane at each surface so that the outermost bond
of each surface turns ferromagnetic. Te is gray; planes are
labeled Te$_1$, Mn$_1$, Te$_2$,\dots{} from the surface, the numbering used
in Figs.~\ref{fig:exchange} and~\ref{fig:bands}. (b)~$\Delta
E=E_{\mathrm{recon}}-E_{\mathrm{A}}$ per $1\times1$ surface
(negative: the reconstruction is the ground state) against slab thickness
($6$--$10$~Mn layers), each state in its own geometry, relaxed without
spin--orbit coupling and evaluated with it. The enumeration behind these energies is described in
Ref.~\cite{SM}.}
\label{fig:recon}
\end{figure}

\begin{figure}
\includegraphics[width=\linewidth]{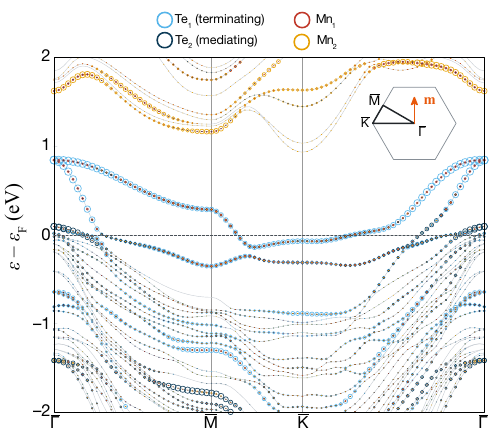}
\caption{Band structure of the reconstructed slab along
$\bar\Gamma$--$\bar{\mathrm{M}}$--$\bar{\mathrm{K}}$--$\bar\Gamma$, for the
same one-face-passivated slab as Fig.~\ref{fig:fs}.
Open-circle radius gives each state's weight on each of the outermost four planes of the metallic face, one color per plane (Te$_1$, Mn$_1$, Te$_2$, Mn$_2$ of
Fig.~\ref{fig:recon}(a)). Inset: the surface Brillouin zone with the sampled path
(heavy line); the arrow $\mathbf{m}$ marks the common direction of the two surface Mn
moments.}
\label{fig:bands}
\end{figure}

\textit{Surface magnetic reconstruction}---Enumerating all $32$ collinear
spin configurations of the 6-Mn slab returns a ground state that is not the
bulk-continued A-type stacking
but the one obtained from it by reversing the spins of the outermost Mn plane
at each surface. The two outermost Mn planes of each surface are then ferromagnetic
while the interior is unchanged, every moment lying in the basal
plane [Fig.~\ref{fig:recon}(a)]. With each state relaxed in its own
geometry, the reconstruction energy $\Delta E\equiv
E_{\mathrm{recon}}-E_{\mathrm{A}}$, the total energy of the reconstructed slab
minus that of the A-type slab, is $-17.1$~meV per $1\times1$
surface [Fig.~\ref{fig:recon}(b)]. The
reconstruction survives in the thicker $8$- and $10$-Mn slabs and across
$\Ueff=2$--$5$~eV on the outermost Mn plane~\cite{SM}.

\begin{figure}
\includegraphics[width=\linewidth]{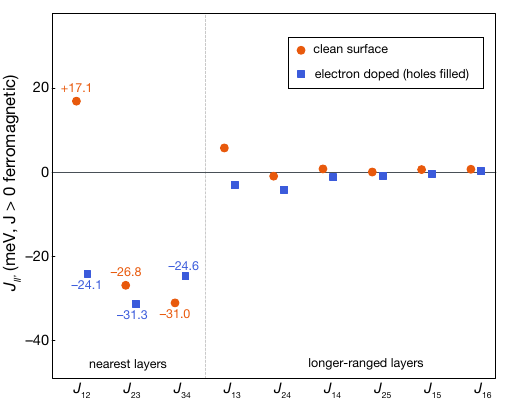}
\caption{Layer-resolved interlayer exchange, with the value in meV printed
at the nearest-layer points. Circles: the clean surface, which carries one ligand hole per
$1\times1$ surface. Squares: the electron-doped slab, with the surface holes
filled by one added electron per $1\times1$ surface. Both series are at the
same frozen geometry.
$J_{ll'}$ is labeled by the Mn-plane pair it couples, in the numbering of
Fig.~\ref{fig:recon}(a) continued into the slab; the mirror-equivalent pair is
implied.}
\label{fig:exchange}
\end{figure}

The same reversal appears in the exchange constants. Fitting the enumeration to the layer model gives a ferromagnetic
$J_{12}=+17.1$~meV against antiferromagnetic subsurface and
interior bonds ($J_{23}=-26.8$, $J_{34}=-31.0$~meV)
[Fig.~\ref{fig:exchange}, circles]. (Note that the agreement of $J_{12}$ with the
reconstruction energy per surface is a numerical coincidence: the latter is a
total-energy difference between relaxed slabs, not an exchange constant.) This model orders collinear stackings and nothing else. We also computed the spin-wave spectrum of the reconstructed surface: it is
stable everywhere in the surface Brillouin zone, and two surface magnon
branches split off above the rest of the spectrum~\cite{SM}.

\textit{The ligand hole and the surface bands}---Figure~\ref{fig:bands} shows
the layer-resolved band structure of the reconstructed slab. The slab is
metallic: surface bands from the Te dangling bonds cross the
Fermi level $\varepsilon_\mathrm{F}$ while the interior keeps the bulk gap, consistent with
photoemission that places the bulk valence-band maximum $\sim\!80$~meV below
$\varepsilon_\mathrm{F}$~\cite{Zhou2026}. Every band that crosses
$\varepsilon_\mathrm{F}$ is localized on the
outermost planes of the metallic face. The states within $50$~meV of
$\varepsilon_\mathrm{F}$ carry $83\%$ of
their weight on Te and $17\%$ on Mn. The hole therefore sits on the anion, and the Mn remain
Mn$^{2+}$. Two Te sites play distinct roles. With the holes on each ion counted as
its integrated weight in the unoccupied valence states, the
terminating Te carries the largest share ($0.51$ of the one hole per surface)
but mediates no interlayer
bond. The Te between the two reversing Mn planes, the mediator of
$J_{12}$, still holds $0.21$ holes. Down the bond ladder the mediating-Te hole count falls
to $0.046$ for $J_{23}$ and $0.027$ for $J_{34}$, leaving the
$J_{12}$ mediator the most hole-rich Te inside the
slab. Of the nearest-layer bonds, only the one whose
mediator carries the hole reverses [Fig.~\ref{fig:exchange}, circles]. With the outermost Mn bilayer ferromagnetic, the surface is left with an
uncompensated moment relative to the bulk-continued termination.

If these holes drive the $J_{12}$
reversal, removing them should restore the bulk order. We do it electronically,
adding one electron per $1\times1$ surface against a uniform
compensating background, at frozen geometry and frozen chemistry, and
recompute the interlayer configurations there. The reconstruction energy changes sign: with the
holes filled it is $+43.8$~meV per surface, and the
bulk-continued stacking is the ground state again. The profile of
Fig.~\ref{fig:exchange} (squares) is almost entirely antiferromagnetic at that endpoint, and $J_{12}$
has moved by $-41.2$~meV against at most $8.9$~meV for any other constant: the
carrier acts on the one bond whose mediator carries it. Two non-energetic observables confirm that the added charge
reaches the surface: the terminating-Te moment vanishes, and the surface Mn
moment recovers to its interior value.

The ferromagnetic sign of $J_{12}$ follows from how the hole is filled. The hole on the mediating
Te has to be filled, and only a Mn can fill it. A half-filled $d^5$ shell has no minority electron to
give, so both Mn of the bond can feed the one hole only if their moments are
parallel. The bond is therefore ferromagnetic through this
channel~\cite{Zener1951b,AndersonHasegawa1955}. Filling the anion shell
closes that channel, and the bond returns to the antiferromagnetic coupling of
the bulk~\cite{Mazin2023}. The two channels enter with opposite sign, so removing the
hole does not merely weaken the bond but reverses it. This is the
hole-mediated exchange of dilute magnetic
semiconductors~\cite{Story1986,Ferrand2001,Kepa2003} at the opposite extreme,
with a moment on every cation and a hole, overwhelmingly Te $5p$, in every
surface cell. \aMnTe{} itself agrees---Li doping makes
the crystal more strongly hole-doped, and inelastic neutron scattering
finds its antiferromagnetic nearest-neighbor interlayer exchange $8\%$ weaker
than in the undoped crystal~\cite{Zhang2025}.

\textit{Constant-energy contours}---Figure~\ref{fig:fs} shows the
constant-energy contours of the reconstructed surface at
$\varepsilon_\mathrm{F}$ and at $0.1$ and $0.2$~eV below it. The contours at
$\varepsilon_\mathrm{F}$ test the surface metal against photoemission; the
deeper ones carry the feature that distinguishes the two stackings in the
data. At
$\varepsilon_\mathrm{F}$ the surface puts two small
pockets at $\bar\Gamma$ and, around them, two larger sheets, all bright: their
weight sits in the outermost few planes rather than spreading through the
slab. The two $\bar\Gamma$ pockets
are the counterparts of the two states that surface-sensitive photoemission
resolves crossing $\varepsilon_\mathrm{F}$, the states that experiment
labels $\alpha$ and $\alpha_1$~\cite{Zhou2026}---a correspondence with no adjustable quantity,
the momentum scale being absolute on both sides and the Fermi level
unshifted. This correspondence tests the surface metal, not the stacking:
the bulk-continued slab places pockets of the same kind at
$\bar\Gamma$~\cite{SM,Zhou2026}.
At $\varepsilon_\mathrm{F}$ every sheet the surface carries is of this kind
(Fig.~\ref{fig:bands}). Sheets of interior character appear only below
$\varepsilon_\mathrm{F}$,
seen as the darker sections of the deeper panels of Fig.~\ref{fig:fs}: the
same crossover the experiment reports~\cite{Zhou2026}.

The six ellipses that surface-sensitive photoemission resolves
at binding energies of $0.1$--$0.2$~eV, $\beta_1$ in the notation of
Ref.~\cite{Zhou2026}, lie with their long axes along
$\bar\Gamma$--$\bar{\mathrm{K}}$. At those energies the reconstruction puts
its surface petals along $\bar\Gamma$--$\bar{\mathrm{K}}$ (the sheet marked
by arrows in Fig.~\ref{fig:fs}), on the measured azimuth; the A-type
surface puts them along
$\bar\Gamma$--$\bar{\mathrm{M}}$~\cite{SM}. But bulk bands can claim the
same azimuth: bulk $k_z=0$ intensity lies along
$\bar\Gamma$--$\bar{\mathrm{K}}$ too~\cite{Krempasky2024}, and
Ref.~\cite{Zhou2026} assigns $\beta_1$ to bulk bands by a calculation
that assumes the bulk-continued stacking, not by a measurement---a tenable
reading, and one under which $\beta_1$ says nothing about the surface
order. If $\beta_1$ is instead a surface state, the azimuth decides between
the stackings. Surface states do not disperse with $k_z$ while bulk bands
do: a photon-energy scan across $\beta_1$ would read out its character
directly.

\textit{Conclusion}---The (0001) termination of \aMnTe{} leaves one
ligand hole per cell of the sharp $1\times1$ surface seen in
diffraction~\cite{Martuza2025}, and the hole reverses the one interlayer
coupling whose mediating Te carries it: the surface ground state is a
magnetically reconstructed stacking with a ferromagnetic outermost Mn bilayer,
$17$--$18$~meV per $1\times1$ surface below the bulk-continued order in every
slab we computed, from six to ten Mn layers. The
coupling is written by the carriers and can be rewritten with them: refilling the
hole electronically restores the bulk order. Chemical doping, surface
chemistry, or an electrostatic gate all act on the same knob, the surface
carrier density. It tunes the strength of the coupling and, by reversing its
sign, selects which magnetic order the surface adopts---the gate being a route to
electric-field control of
magnetism~\cite{Ohno2000,Weisheit2007,Heron2014,Matsukura2015,Taniyama2024},
here at the surface of an altermagnet.

The interior keeps its altermagnetic stacking; only within the reconstructed
bilayer are the two spin sublattices, related in the bulk by a rotation,
parallel rather than antiparallel. The surface
magnetic order becomes an input to theoretical analysis of surface-
or interface-derived
transport~\cite{Zhou2026,SurfaceAHE2026,Benny2026,Sattigeri2023}, which has
so far had every reason to build on the bulk-continued order. The
reconstruction also gives the compensated
altermagnet an uncompensated moment at each surface, exchange-coupled to the
N\'eel vector, confined to the termination, and switched off with the hole.
Scanning-probe magnetometry already places the magnetization of \aMnTe{}
films at the surface, but out of plane and through a capping
layer~\cite{Du2026}; the in-plane moment of the clean termination is
untested. The measurements are within reach. The contours computed for the
reconstructed
surface already reproduce the Fermi-level pockets that surface-sensitive
photoemission resolves~\cite{Zhou2026}, a check on the surface metal rather
than on the stacking; a photon-energy
scan across $\beta_1$, or magnetometry on an uncapped film, would
test the reconstruction directly.

\textit{Acknowledgments}---This work was partly supported by JSPS KAKENHI
Grant Number JP24K01144
and MEXT-DXMag Grant Number JPMXP1122715503. The calculations were
partly carried out by using facilities of the Supercomputer Center at the
Institute for Solid State Physics, the University of Tokyo, and
TSUBAME4.0 supercomputer at Institute of Science Tokyo.

\textit{Data availability}---The data supporting the findings of this study
are available from the first author, T.T., upon reasonable request.

\end{document}

\title{Supplemental Material for\\
``Magnetic reconstruction of the altermagnet
\aMnTe(0001) surface driven by ligand holes''}

\author{Tomonori Tanaka}
\author{Yoshihiro Gohda}
\affiliation{Department of Materials Science and Engineering, Institute of Science Tokyo, Yokohama 226-8501, Japan}

\date{\today}

\maketitle

\section{Enumeration and layer-resolved exchange fit}\label{sm:fit}

The settings common to every campaign are those of the Methods section of the
main text; each
section below states where it departs from them. The slabs are
Mn$_{N}$Te$_{N+1}$ stacks, so the $13$-, $17$- and $21$-plane slabs carry
$N=6$, $8$ and $10$ Mn planes.
No cell is optimized in any slab: relaxations move ions at fixed cell until forces fall below
$0.01$~eV/\AA{}. Local moments and hole shares are projections onto the atom-centered
augmentation spheres, whose radius is $1.217$~\AA{} for both
species.

\begin{figure}
\includegraphics[width=0.56\textwidth]{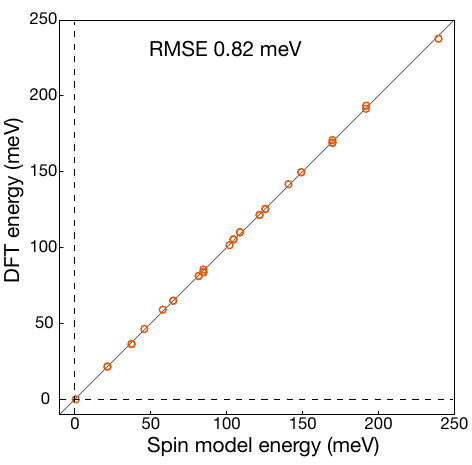}
\caption{Model against density functional theory (DFT) for the fit used
throughout (spin--orbit coupled,
relaxed geometry, one hole per $1\times1$ surface).
Energies of all $32$ collinear interlayer configurations from the layer model
of Eq.~(\ref{eq:heis}), against the DFT energies they are fitted to, both
referred to the DFT ground-state energy. All $32$ are plotted; some points
overlap through the mirror degeneracy.}
\label{fig:parity}
\end{figure}

We computed \emph{all} inequivalent
collinear interlayer
spin configurations of the 6-Mn slab, $2^{N-1}=32$ after removing the global
flip. The geometry was relaxed within an assumed magnetic state, the enumeration
was rerun on the resulting geometry, and the cycle was repeated until the
lowest configuration coincided with the state the geometry had been relaxed
in.
Each configuration is an unconstrained self-consistent solution: the collinear
pattern sets the starting moments and no penalty holds them there afterwards,
and we verified that all $32$ converge to the pattern they were started in.
The energies are fitted to
\begin{equation}
E(\{\sigma_l\}) = E_0 - \sum_{l<l'} J_{ll'}\,\sigma_l\sigma_{l'},
\label{eq:heis}
\end{equation}
with $\sigma_l=\pm1$ the orientation of the (ferromagnetic) $l$th Mn plane,
$E_0$ a configuration-independent constant, and $J>0$ favoring parallel
(ferromagnetic) alignment. The slab mirror symmetry
ties the $J_{ll'}$ into nine classes. The model reproduces the $32$ energies to a
root-mean-square error of
$0.82$~meV against a configuration ladder $238$~meV wide
[Fig.~\ref{fig:parity}].

\section{Robustness against the surface Hubbard $\Ueff$}\label{sm:robust}

Because the metallic surface could
warrant a different on-site Coulomb correction $\Ueff$ on the undercoordinated Mn, the outermost Mn plane
of each face (Mn$_1$ and Mn$_6$) was given its own value $\Usurf$, the four
interior planes staying at $4$~eV, and the A-type (bulk antiferromagnetic)
and reconstructed states recomputed for
$\Usurf=2$, $3$, $4$ and $5$~eV. Repeating the relaxation at every
$\Usurf$ tests the geometry as well as the energetics.
The reconstruction remains the ground state throughout: at
$\Usurf=2$, $3$, $4$ and $5$~eV it lies $-9.8$, $-14.4$, $-17.1$ and
$-19.1$~meV per surface below the A-type stacking. Its margin grows
with $\Usurf$ but never approaches zero over a range that brackets
any defensible choice.

\section{Stability of the Te termination}\label{sm:term}

\begin{figure}
\includegraphics[width=0.62\textwidth]{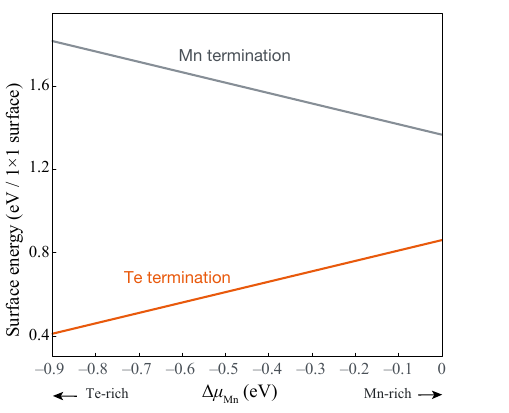}
\caption{Surface energy of the Te- and Mn-terminated
MnTe(0001) surfaces across the accessible chemical-potential window
($\Delta H_f=-0.9$~eV shown). Each termination is drawn in the lower of the two
stackings computed for it: the reconstructed stacking for Te, A-type for Mn.}
\label{fig:gamma_term}
\end{figure}

The Mn- and Te-terminated slabs
have different compositions, so their energies cannot be compared
directly; the quantity that can be is the surface energy
[Fig.~\ref{fig:gamma_term}], the
excess energy of the slab over the same atoms drawn from reservoirs. For a
symmetric slab with $N_{\mathrm{Mn}}$ Mn and $N_{\mathrm{Te}}$ Te atoms and
two equivalent $1\times1$ surfaces,
\begin{equation}
\gamma=\frac{1}{2}\bigl(E_{\mathrm{slab}}
 -N_{\mathrm{Mn}}\,\mu_{\mathrm{Mn}}
 -N_{\mathrm{Te}}\,\mu_{\mathrm{Te}}\bigr),
\label{eq:gamma_def}
\end{equation}
so that $\gamma$ is an energy per $1\times1$ surface, the unit used
throughout, where $E_{\mathrm{slab}}$ is the slab total energy and $\mu_{\mathrm{Mn}}$,
$\mu_{\mathrm{Te}}$ the chemical potentials of the reservoirs the surface
exchanges atoms with. The two potentials are not independent: the surface
coexists with the bulk crystal it terminates, so transferring a formula unit
between slab and bulk must cost nothing,
$\mu_{\mathrm{Mn}}+\mu_{\mathrm{Te}}=E_{\mathrm{MnTe}}$, with
$E_{\mathrm{MnTe}}$ the bulk \aMnTe{} energy per formula unit. Using this to
eliminate $\mu_{\mathrm{Te}}$ from Eq.~(\ref{eq:gamma_def}),
\begin{equation}
\gamma(\mu_{\mathrm{Mn}})=\frac{1}{2}\Bigl[E_{\mathrm{slab}}
 -N_{\mathrm{Te}}\,E_{\mathrm{MnTe}}
 -\bigl(N_{\mathrm{Mn}}-N_{\mathrm{Te}}\bigr)\,\mu_{\mathrm{Mn}}\Bigr].
\label{eq:gamma}
\end{equation}
The remaining variable $\mu_{\mathrm{Mn}}$ ranges over a finite window. If it
rose above the energy per atom of elemental Mn, $E_{\mathrm{Mn}}$, that phase
would precipitate; if
$\mu_{\mathrm{Te}}$ rose above the trigonal-Te energy per atom
$E_{\mathrm{Te}}$ (computed in the same setup), elemental Te would. The two
bounds are separated by the formation enthalpy
$\Delta H_f=E_{\mathrm{MnTe}}-E_{\mathrm{Mn}}-E_{\mathrm{Te}}$. Measure
$\mu_{\mathrm{Mn}}$ from the first of those bounds,
$\Delta\mu_{\mathrm{Mn}}\equiv\mu_{\mathrm{Mn}}-E_{\mathrm{Mn}}$, and use the
formation enthalpy to eliminate the elemental-Mn energy it refers to,
$E_{\mathrm{Mn}}=E_{\mathrm{MnTe}}-E_{\mathrm{Te}}-\Delta H_f$; then
\begin{equation}
\mu_{\mathrm{Mn}}=E_{\mathrm{MnTe}}-E_{\mathrm{Te}}-\Delta H_f
+\Delta\mu_{\mathrm{Mn}},
\label{eq:mu_shift}
\end{equation}
a substitution that removes $E_{\mathrm{Mn}}$ from the problem: the
elemental-Mn reference now enters only through $\Delta H_f$.
It puts the physical window at
$\Delta H_f\le\Delta\mu_{\mathrm{Mn}}\le0$: precipitation of bulk Mn at
$\Delta\mu_{\mathrm{Mn}}=0$ (Mn-rich), of trigonal Te at
$\Delta\mu_{\mathrm{Mn}}=\Delta H_f$, i.e.\
$\mu_{\mathrm{Te}}=E_{\mathrm{Te}}$ (Te-rich).
Substituting into Eq.~(\ref{eq:gamma}) gives the form actually evaluated,
\begin{equation}
\gamma(\Delta\mu_{\mathrm{Mn}})=\frac{1}{2}\Bigl[E_{\mathrm{slab}}
 -N_{\mathrm{Te}}\,E_{\mathrm{MnTe}}
 -\bigl(N_{\mathrm{Mn}}-N_{\mathrm{Te}}\bigr)
  \bigl(E_{\mathrm{MnTe}}-E_{\mathrm{Te}}-\Delta H_f
  +\Delta\mu_{\mathrm{Mn}}\bigr)\Bigr],
\label{eq:gamma_dmu}
\end{equation}
in which the only quantities computed from first principles are the slab
energy, bulk \aMnTe{} and trigonal Te. The
formation enthalpy enters only as the window width; we plot
$\Delta H_f=-0.9$~eV.

Treating $\Delta H_f$ as a parameter rather than computing it avoids the one badly
posed reference in the problem: $\Ueff=4$~eV is chosen for the localized $d^5$
shell of Mn$^{2+}$, and applying it unchanged to the itinerant $d$ band of
elemental $\alpha$-Mn would bias $\Delta H_f$ by an amount we cannot
control. This is the incomplete error cancellation that makes computed
formation enthalpies of transition-metal compounds unreliable, and which is
normally repaired by correcting the elemental references themselves---by
fitting them to measured formation enthalpies~\cite{Stevanovic2012}, or by
fitting one correction per anion~\cite{Wang2006}. Here the elemental
reference is removed from the calculation altogether instead, at the price of
a single parameter whose only role is to set the width of the window. No
elemental Mn calculation enters this work, and at the Te-rich
edge none is needed even in principle: equilibrium with trigonal Te fixes
$\mu_{\mathrm{Te}}=E_{\mathrm{Te}}$ and hence
$\mu_{\mathrm{Mn}}=E_{\mathrm{MnTe}}-E_{\mathrm{Te}}$, so $\Delta H_f$
cancels identically from Eq.~(\ref{eq:gamma}) and the margin quoted at
that edge is independent of it.
Both terminations were computed at matched thickness (15 planes, $1\times1$,
sampled on an $8\times8\times1$ $k$-point mesh), and both references with the
same settings,
including spin--orbit coupling.

Of the two, the Te termination is the stable one across the
entire window. At the Te-rich edge, the
$\Delta H_f$-free point, $\gamma=0.41$~eV per surface against
$1.82$~eV for
the Mn termination, a margin of $1.41$~eV; at the Mn-rich edge the two are
$0.86$ and $1.37$~eV, a margin
of $0.51$~eV. The Mn termination is therefore never competitive. Only the
Mn-rich margin depends on the formation enthalpy, and linearly: it is
$1.41$~eV$\,-|\Delta H_f|$, so the ordering survives any
$|\Delta H_f|<1.41$~eV.

The magnetic state in which each termination is evaluated is not assumed but
determined by enumeration, on thinner, $11$-plane symmetric slabs: all
$2^{N-1}$ collinear interlayer configurations, each relaxed in its own
geometry---$32$ for the Mn-terminated slab, which carries six Mn planes, and
$16$ for the Te-terminated one, which carries five. The Te
termination comes out at the reconstructed stacking, the same ordering the main
text finds; the Mn termination comes out at the bulk-continued A-type one, with
its reconstruction analogue $213$~meV per slab higher. The curves above are the
spin--orbit calculations for those two states at the matched thickness.

\section{The one-face hydrogen passivation}\label{sm:hpass}

\begin{figure}
\includegraphics[width=\textwidth]{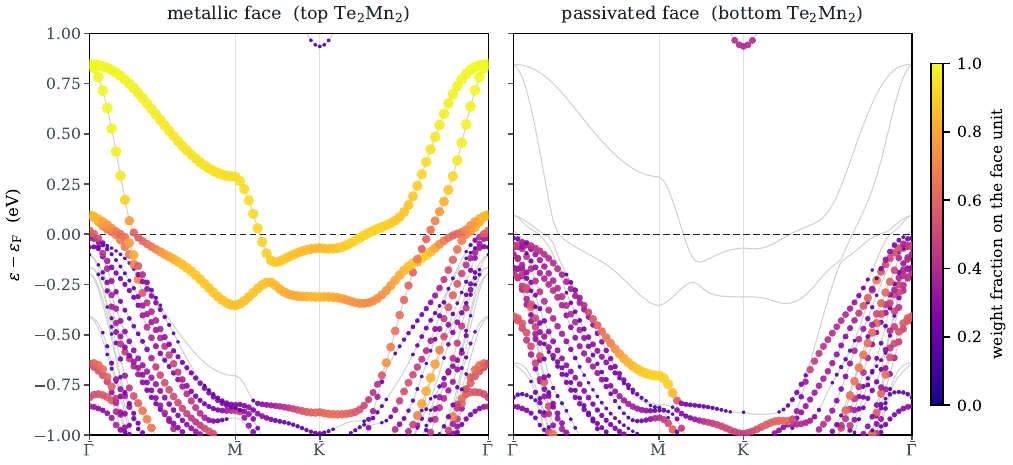}
\caption{The bands of the one-face-passivated reconstructed slab along
$\bar\Gamma$--$\bar{\mathrm{M}}$--$\bar{\mathrm{K}}$--$\bar\Gamma$, drawn
twice: marker area and color give the state's weight fraction on the
outermost four planes of the metallic face (left) and of the passivated face
(right, its pseudo-hydrogens not counted).}
\label{fig:hpass}
\end{figure}

A symmetric slab carries two equivalent metallic faces
whose surface states hybridize through the slab. The total energies barely
register it---the reconstruction energy moves by $0.8$~meV per surface
between six and ten Mn planes [Fig.~\ref{M-fig:recon}(b) of the main
text]---but every surface band is split in two by an amount that is an
artifact of the finite thickness. For the band structure and the constant-energy contours of the
main text (Figs.~\ref{M-fig:bands} and \ref{M-fig:fs}) one face was therefore
passivated: three pseudo-hydrogens of valence charge $0.33\,e$, one along
each of the three Te--Mn bonds that the cut severs, cap the terminating Te
of that face. They are the only atoms free
to move; the rest of the slab is held at the geometry it relaxed to with the
clean face, and the cell carries a dipole
correction~\cite{Neugebauer1992,Bengtsson1999} for
the now asymmetric slab. Between them they supply the charge that fills
that face's dangling-bond hole, so the cell stays neutral and the slab
keeps a single metallic face carrying one hole. All other settings are
unchanged, except that the vacuum is increased to $20$~\AA{}. The constant-energy
contours are non-self-consistent eigenvalues on a $\Gamma$-centered
$36\times36\times1$ grid, contoured after interpolation onto a $501\times501$
mesh.

One third of an electron on each of the three severed
bonds saturates them and gaps that face at the Fermi level
$\varepsilon_\mathrm{F}$
[Fig.~\ref{fig:hpass}].
The construction also isolates the reconstruction
on a single surface. Reversing the outermost Mn of the metallic face
alone, with the passivated face left bulk-continued in both runs and with the
geometry, the cell and the electron count held fixed, lowers the energy by
$23.0$~meV per surface. The reconstruction is therefore a property of the
hole-bearing face by itself: it does not need the second surface, and the
hole was removed from the other face chemically, not by a
compensating background.

\section{Constant-energy contours: the full zone and the A-type stacking}\label{sm:fsbz}

\begin{figure}
\includegraphics[width=\textwidth]{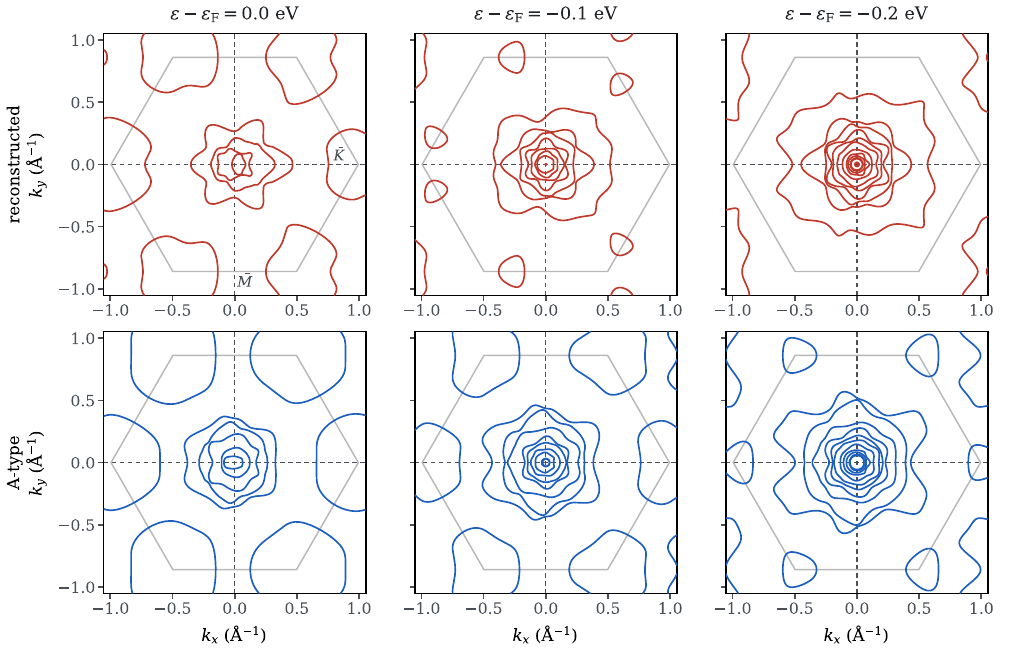}
\caption{Constant-energy contours of the two one-face-passivated slabs over
the full surface Brillouin zone (gray hexagon), one row per magnetic state:
reconstructed above, bulk-continued A-type below, at
$\varepsilon-\varepsilon_\mathrm{F}=0$, $-0.1$, $-0.2$~eV.}
\label{fig:fsbz}
\end{figure}

Figure~\ref{fig:fsbz} draws the contours of both slabs over the whole surface
Brillouin zone. The window of Fig.~\ref{M-fig:fs} of the main text,
$|k_{x,y}|<0.6$~\AA$^{-1}$ ($k_x$ along
$\bar\Gamma$--$\bar{\mathrm{K}}$, $k_y$ along
$\bar\Gamma$--$\bar{\mathrm{M}}$), is about the extent of the published
constant-energy maps~\cite{Zhou2026S}; what it
clips is the six $\bar{\mathrm{K}}$ corner pockets, present for both
magnetic states and the most strongly surface-localized states anywhere in
the zone; the measurement of Ref.~\cite{Zhou2026S} does
not reach them. Apart from the corners, and the outer tips of the
$-0.2$~eV petals along $\pm k_x$, the zone outside the window is empty.
The lower row is the bulk-continued A-type slab, shown for reference at its
own $\varepsilon_\mathrm{F}$. It too carries the two small $\bar\Gamma$
pockets at $\varepsilon_\mathrm{F}$: those pockets belong to the Te
termination, not to the stacking. Its petal-bearing outermost sheet is as
surface-localized as the reconstructed ones: its weight fraction on the
outermost four planes of the metallic face, in the sense of
Fig.~\ref{M-fig:fs} of the main text, is $0.72$--$0.87$ over the three
energies, against $0.55$--$0.76$ for the sheets that carry the
reconstructed petals. What differs is the orientation: at
$\varepsilon_\mathrm{F}$ the petals of the outermost sheet point
along $\bar\Gamma$--$\bar{\mathrm{K}}$ on the reconstructed surface and along
$\bar\Gamma$--$\bar{\mathrm{M}}$ on the A-type one; at $-0.1$~eV the outermost
reconstructed sheet has rounded, and the sheet inside it carries the same
$\bar\Gamma$--$\bar{\mathrm{K}}$ petals. Quantitatively, for each contour
of Fig.~\ref{fig:fsbz} we read off its largest radius from $\bar\Gamma$
within $\pm4^\circ$ of the six $\bar\Gamma$--$\bar{\mathrm{K}}$ azimuths
and, separately, within $\pm4^\circ$ of the six
$\bar\Gamma$--$\bar{\mathrm{M}}$ ones. The ratio of the former to the
latter, for the petal-bearing sheet of each surface, is $1.4$ at
$\varepsilon_\mathrm{F}$ and $1.3$ at $-0.1$ and $-0.2$~eV on the
reconstructed side, against $0.90$, $0.81$ and $0.74$ on the A-type side,
where the petals stay on the outermost sheet throughout. Widening the
windows to $\pm10^\circ$, or replacing the largest radius within a family
by its mean, moves none of these ratios across unity; and no sheet of the
A-type slab, at any of the three energies, exceeds a ratio of $1.06$.

With spin--orbit coupling and the N\'eel axis along
$\langle 1\bar{1}00\rangle$---the $+k_y$ direction of Fig.~\ref{fig:fsbz}, and
the easy axis at which the bulk bands of \aMnTe{} have been calculated and
measured~\cite{Krempasky2024S}---one point operation survives in the slab, and
it is antiunitary: time reversal
combined with the vertical mirror whose plane contains that axis, so the
magnetic point group is $m'$. This reduction of an altermagnet's symmetry by
a surface, and what it leaves
of the spin splitting, are treated generally in Ref.~\cite{Sorn2026S}. Because
time reversal also sends $\mathbf{k}\to-\mathbf{k}$, what is left acts in the
zone as a reflection about the axis perpendicular to the N\'eel direction: in
the frame drawn here the contours are even across $k_y=0$ and asymmetric across
$k_x=0$.

\section{Spin waves of the reconstructed surface}\label{sm:magnon}

The layer model of Eq.~(\ref{eq:heis}) only orders collinear stackings. It
cannot say whether the reconstructed state survives small noncollinear
deviations at finite in-plane wavevector, and it carries no in-plane exchange
at all, since in a $1\times1$ cell every in-plane neighbor is a periodic image
of the same atom.

\begin{figure}[b]
\includegraphics[width=0.72\textwidth]{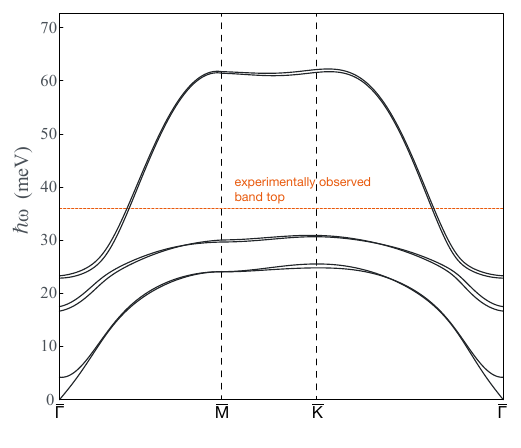}
\caption{Linear spin-wave dispersion of the reconstructed slab along
$\bar\Gamma$--$\bar{\mathrm{M}}$--$\bar{\mathrm{K}}$--$\bar\Gamma$ of the $1\times1$ surface
Brillouin zone, from the spin-cluster expansion fitted to constrained
noncollinear calculations. The dashed line is the measured top of the bulk
\aMnTe{} magnon band, $\approx36$~meV~\cite{Liu2024S}.}
\label{fig:magnon}
\end{figure}
A spin-cluster expansion~\cite{Drautz2004} was fitted with
Magesty.jl~\cite{Tanaka2026}
to constrained-direction noncollinear DFT with spin--orbit coupling on a
$3\times3$ in-plane supercell of the relaxed slab ($117$ atoms; the
$3\times3$ is what makes in-plane exchange visible). This campaign departs
from the common settings in three ways, all of them concessions to the
supercell: it samples the Brillouin zone on a $\Gamma$-centered
$3\times3\times1$ $k$-point mesh, $9\times9\times1$ per $1\times1$ cell,
converges total
energies to $10^{-5}$~eV, and carries $10$~\AA{} of vacuum. The model fitted is
\begin{equation}
\mathcal{H}=-\sum_{i<j}J_{ij}\,\mathbf{S}_i\cdot\mathbf{S}_j
+\sum_{i<j}\mathbf{D}_{ij}\cdot(\mathbf{S}_i\times\mathbf{S}_j)
+\sum_{i<j}\mathbf{S}_i\cdot\mathsf{\Gamma}_{ij}\cdot\mathbf{S}_j
+\sum_i\mathbf{S}_i\cdot\mathsf{A}_i\cdot\mathbf{S}_i,
\label{eq:sce}
\end{equation}
with $|\mathbf{S}|=5/2$: $J_{ij}$ is the isotropic exchange, positive for
ferromagnetic as in Eq.~(\ref{eq:heis}); $\mathbf{D}_{ij}$ is the
Dzyaloshinskii--Moriya vector and $\mathsf{\Gamma}_{ij}$ the symmetric
traceless part of the same pair tensor; and $\mathsf{A}_i$ is the on-site
anisotropy. Magnitudes are per Mn pair and are not to be compared with the
per-plane-pair constants of Fig.~\ref{M-fig:exchange} of the main text.

The training configurations are drawn from the mean-field distribution of a
classical Heisenberg model~\cite{Tanaka2026} at a scaled temperature
$\tau=T/T_c$, with $T_c$
the mean-field ordering temperature. Each moment direction $\hat{\mathbf{e}}_i$ is
drawn independently about its ground-state direction
$\hat{\mathbf{e}}_i^{\,0}$ at fixed magnitude, from a von Mises--Fisher distribution
whose concentration follows the mean-field magnetization $m$ at that
temperature:
\begin{equation}
p(\hat{\mathbf{e}}_i)=\frac{\kappa}{4\pi\sinh\kappa}\,
e^{\,\kappa\,\hat{\mathbf{e}}_i\cdot\hat{\mathbf{e}}_i^{\,0}},
\qquad \kappa=\frac{3m}{\tau},
\qquad m=\coth\frac{3m}{\tau}-\frac{\tau}{3m}.
\label{eq:vmf}
\end{equation}
We drew $100$ configurations at $\tau=0.05$, deep in the ordered phase
($m=0.983$, $\kappa=59$). The training set therefore
samples the
immediate neighborhood of the ground state, where the transverse fluctuations
are very small---the regime linear spin-wave theory itself expands in. Each
configuration is also given a random rigid rotation as a whole: turning the
spin arrangement bodily against the lattice is what determines the anisotropic
terms, the on-site $\mathsf{A}_i$ in particular. The $147$
symmetry-adapted coefficients of Eq.~(\ref{eq:sce}) were fitted to the
constraining torques, the energies fixing only the constant offset; the model
reproduces the DFT energies with $R^{2}=0.996$ and a root-mean-square error of
$3.3$~meV per $117$-atom cell, against a configuration spread of $206$~meV.

The spin waves of that model were then computed in linear spin-wave
theory, as implemented in Sunny.jl~\cite{Sunny}.
On a $36\times36$ grid covering the surface Brillouin zone
no mode is negative ($0$ of $1296$ wavevectors): the only zero is the
Goldstone mode at $\bar\Gamma$ of a rigid in-plane rotation---exactly
gapless here, since the threefold surface symmetry forbids a rank-two
in-plane anisotropy---and the lowest finite-wavevector mode lies at
$2.07$~meV. The reconstruction is therefore a local minimum against
all small-amplitude, finite-wavevector deviations, not merely the lowest of
the $32$ collinear stackings.

Two branches split off above the rest and reach $62.2$~meV along this path,
the other four staying below $31$~meV [Fig.~\ref{fig:magnon}].
The interior of the slab carries nothing at $62$~meV, so the two upper
branches are
surface excitations, expelled above the bulk band by a surface whose exchange
is the stronger. The two branches are one mode per face rather than two
distinct excitations: the two surfaces of the slab are equivalent, so the surface mode
appears twice, as the even and odd combinations of the two faces, split by
$0.3$--$1.1$~meV along this path through the six Mn planes between them. That
splitting is an artifact of the finite thickness; in a thick film the faces
decouple and the two merge into one surface branch.

The two surface branches are lifted by the in-plane nearest-neighbor exchange
of the outermost
Mn plane, ferromagnetic at $J_{ij}=+2.3$~meV against $-0.1$~meV within the
interior planes. The interior is the bulk situation: in bulk
\aMnTe{} the in-plane exchange is antiferromagnetic and it is the interlayer
bonds that hold a plane ferromagnetic~\cite{Mazin2023S}. Replacing the whole in-plane exchange of the
outermost planes by that of the interior ones removes both
branches, and the spectrum falls below $31.8$~meV.